\documentclass[reprint,secnumarabic,amssymb, nobibnotes, aps, pre]{revtex4-1}

\newcommand{\jgr}{J. Geophys. Res.~}
\newcommand{\apjl}{Astrophys. J. Lett.~}

\usepackage{graphicx}
\usepackage{dcolumn}
\usepackage{bm}
\usepackage{hyperref}
\usepackage[mathlines]{lineno}
\usepackage{amsmath,amsthm,amssymb,bm}

\begin{document}

\preprint{APS/123-QED}

\title{Magnetic moment non-conservation in magnetohydrodynamic turbulence models}

\author{S.~Dalena$^{1,2}$}
\author{A.~Greco$^{1}$}
\author{A.~F.~Rappazzo$^{2}$}
\author{R.~L. Mace$^{3}$}
\author{W.~H. Matthaeus$^{2}$}

\affiliation{%
$^{1}$Dipartimento di Fisica, Universit\`a della Calabria, I-87036 Cosenza, Italy\\%
$^{2}$Bartol Research Institute, Department of Physics and Astronomy, University of Delaware, Newark, DE 19716, USA\\
$^{3}$School of Physics, University of KwaZulu-Natal, Westville Campus, South Africa}

\date{\today}

\begin{abstract}

The fundamental assumptions of the adiabatic theory do
not apply in presence of sharp field gradients as well as in presence of well developed  
magnetohydrodynamic turbulence. 
For this reason in such conditions the magnetic moment $\mu$ is no longer expected to be constant. 
This can influence particle acceleration and have considerable implications in many 
astrophysical problems. 

Starting with the resonant interaction between ions and a single parallel propagating electromagnetic 
wave, we derive expressions for the magnetic moment trapping width $\Delta \mu$ (defined as the 
half peak-to-peak difference in the particle magnetic moment) and the bounce frequency $\omega_b$.
We perform test-particle simulations to investigate magnetic moment behavior when resonances overlapping occurs and 
during the interaction of a ring-beam particle distribution with a broad-band slab spectrum. 

We find that magnetic moment dynamics is strictly related to pitch angle $\alpha$ for a low level of magnetic 
fluctuation, $\delta B/B_0 = (10^{-3}, \, 10^{-2})$, where $B_0$ is the constant and uniform background magnetic 
field. 
Stochasticity arises for intermediate fluctuation values and its effect on pitch angle is the isotropization of 
the distribution function $f(\alpha)$. This is a transient regime during which magnetic moment 
distribution $f(\mu)$ exhibits a characteristic one-sided long tail and starts to be influenced 
by the onset of spatial parallel diffusion, i.e., the variance $\langle (\Delta z)^2 \rangle$ grows 
linearly in time as in normal diffusion. 
With strong fluctuations $f(\alpha)$  isotropizes completely, spatial diffusion sets in 
and $f(\mu)$ behavior is closely related to the sampling of the varying magnetic field 
 associated with that spatial diffusion. 
\end{abstract}


\maketitle

\section{Introduction}
\label{sec:intro}

In this paper we study magnetic moment $\mu$ conservation for charged particles
in presence of a single electromagnetic wave as well as in presence of turbulent magnetic fields having one dimensional spectra
comparable to those measured in the solar wind.
Magnetic moment conservation is an important topic in plasma physics. 
Indeed, some of the most commonly used theories that describe particle motion in perturbed 
magnetic fields are based on the assumption that 
particles magnetic moment is on average constant over a gyroperiod.
When $\mu$ is not conserved, this approximation is not allowed and its effects can have a bearing
on several astrophysical phenomena such as coronal heating, cosmic ray transport,
temperature anisotropies observed in the solar wind \cite{Marsch91} and 
particle acceleration near reconnection sites \cite{KnizhnikEA11}. 
Furthermore this issue is strictly related to 
particle confinement in plasma machines and dynamically chaotic systems \cite{Chirikov87}. 
Therefore we want to established the validity range of the adiabatic approximation and 
the key mechanisms that regulate magnetic moment non-conservation.

The guiding center approximation \cite{RossiEA70} splits particle motion 
into the motion of the guiding center and the gyromotion around it.
When analyzing charged particle motion in nonuniform electromagnetic fields,
we would like to neglect the rapid and relatively uninteresting gyromotion,
focusing instead on the far slower motion of the guiding center.
Averaging the particle equation of motion over the gyrophase, we obtain 
a reduced equation that describes the guiding center motion.
In the non-relativistic case the equation of motion of the guiding center 
in the direction parallel to the magnetic field reads
\begin{equation}
\frac{dp_{\parallel}}{dt} = -\mu \nabla_{\parallel} B + qE_{\parallel},
\end{equation}
where particle magnetic moment is defined as $\mu = v_{\perp}^2/B$ and 
$\nabla_{\parallel} = (\mathbf{\hat{B}} \cdot \nabla)$ is the spatial derivative along the field direction.
In the perpendicular direction the guiding center drifts with the velocity
\begin{equation}
v_D = \frac{{\bf F} \times {\bf B}}{qB^2},
\end{equation}
where ${\bf F} = [q{\bf E} - \mu \nabla B - (mv^2_{\parallel})\nabla_{\parallel}{\bf B}]$ is the total force acting on the 
guiding center, averaged over a gyroperiod, in the (non-inertial) frame co-moving with the guiding center.
Therefore, as long as a particle moves through slowly varying electric and magnetic fields,
its guiding center behaves like a particle with a magnetic moment $\mu$ conserved.

This approximation is valid when the smallest length-scales of the electromagnetic fields are much larger 
than the particles Larmor radius, i.e., when particle magnetic moment is a constant of motion on average over the particle gyroperiod.
This corresponds to the well-known Born-Oppenheimer approximation in quantum mechanics.
This description for particle motion in a non-uniform magnetic field is also useful for numerical 
simulations. Indeed direct simulations of kinetic equations (Vlasov, Boltzmann) with a large magnetic field require 
the numerical resolution of small spatial and time scales induced by the gyration along the magnetic field.
The guiding center approximation, as well as gyrokinetics, are approximate models describing particle motion 
in presence of a strong magnetic field.
However, the assumption that the scale of variation of the magnetic field is much larger than the particle Larmor radius can break  
down in presence of turbulence. Turbulent magnetic fluctuations are observed in  
space plasmas in practically all environments and at all scales. 
Furthermore the presence of waves in collisionless plasma introduces through wave-particle interactions
a finite dissipation. In this case it seems invalid to resort to a guiding center theory.

When the amplitude of the magnetic fluctuations is lower than that of the mean magnetic field (averaged
over the fluctuations time-scale), a perturbation approach called \emph{quasilinear approximation} is 
applicable \cite{Jokipii66, Urch77, JonesEA98}. 
In this case the \emph{resonant} fluctuations make the dominant contribution to particle scattering.
The resonance condition for wave-particle interaction is given by:
\begin{equation}
\omega - k_{\parallel}v_{\parallel} = n\Omega
\label{eq:res_cond}
\end{equation}
where $\omega$ is the wave frequency, $k_{\parallel}$ and  $v_{\parallel}$ are respectively the wavevector 
and the particle velocity along the mean magnetic field ${\bf B}_0$, and $\Omega = qB/m$ is the particle 
gyrofrequency. Landau resonance \cite{Landau46} is found at $n=0$, while $n= \pm 1$, $\pm 2,\, \ldots$  are the cyclotron 
resonances.
In linear theory these resonances are represented by delta functions. In presence of well-developed 
magnetohydrodynamic turbulence we expect that the discrete resonances to be significantly 
broadened due to the rapid decorrelation of the waves phases in strong turbulence \cite{Chandran00}. 

The particle reaction to the perturbation is always periodic except when 
condition~(\ref{eq:res_cond}) is satisfied. In this case the perpendicular electric force 
due to the wave remains in phase with the particle cyclotron motion and 
particle reaction is \emph{secular} or resonant and, over short times, non-oscillatory. 
The secular electric force acting on a given particle is 
constant over a particle gyroperiod, so that the magnetic moment is no longer conserved. 

Charged particles are scattered by their interaction with the waves and undergo 
pitch angle diffusion.
The pitch angle, $\theta = \arctan(v_{\perp} / v_{\parallel})$, is the angle between 
the direction of the magnetic field and the particle's helical trajectory.
Scattering from magnetic fluctuations causes the distribution of 
pitch angle cosine, $\alpha = v_{\parallel}/|v|$, to become isotropic.
Magnetic moment, $\mu$, is formally related to the time averages of the cosine of  pitch angle by:
\begin{equation} \label{mu}
\mu \sim \frac{v^2_{\perp}}{|B|} =  \frac{v^2}{|B|}(1 - \alpha^2)
\end{equation}
We therefore expect the behavior of magnetic moment to be strongly related to 
pitch angle behavior.

\section{Stochastic motion, trapping width and resonance overlapping}
\label{sec:trapp_theory}

Wave-particle interactions usually involve multiple resonances. Particle motion is substantially
different depending on when these resonances overlap or not.
Numerical simulations show a complex behavior that cannot be approached analytically, e.g., 
it is not possible to write an equation for the evolution of particles distributions 
when two resonances overlap~\cite{SmithEA78}. 
Such motions in the presence of overlapping resonances are commonly labeled \emph{stochastic}.

It is important to distinguish between two different kinds of stochasticity. 
Wave-particle interaction in presence of of uncorrelated small amplitude electromagnetic waves or 
plasma turbulence is called extrinsically diffusive \cite{LichtenbergEA89}.
In this case the regular phase space structure for a 
charged particle interacting resonantly with an electromagnetic wave is perturbed by neighboring uncorrelated 
waves. This leads to \emph{extrinsic stochasticity} and diffusive behavior. 
On the other hand nonlinear systems, such as particle interacting resonantly with a large amplitude 
obliquely propagating (with respect to $\mathbf{B}_0$) electromagnetic plasma wave,  can 
exhibit \emph{intrinsic stochasticity}. Indeed, when the wave amplitude is sufficiently large, the 
resonances at the gyrofrequency harmonics are sufficiently broadened that they overlap 
with adjacent primary resonances. Therefore particles interacting even with a single monochromatic 
wave may exhibit intrinsically stochastic and diffusive behavior~\cite{KarimabadiEA90}. This is the 
regime of  nonlinear diffusion and irreversible chaotic mixing of orbits. 

Because one of the main hypothesis of quasilinear theory is that particles dynamics is adequately modeled by 
their unperturbed trajectories, the quasilinear timescale $\tau_{c}$ must be much smaller than the timescale for the onset of nonlinear orbit effects
$\tau_{nl}$~\cite[cf.][]{Weinstock69, Davidson72}:
\begin{equation}
\tau_{c} \ll \tau_{nl} \sim \frac{1}{\omega_b},
\label{eqn:qlt_orderings}
\end{equation}
where ${\omega_b}$ is the bounce frequency. This means that the turbulent spectrum should be broad enough 
so that the typical timescale for a charged particle to interact with a resonant wave-packet would be much less than 
its typical bounce time, $\tau_b=2\pi/\omega_b$, in a monochromatic wave at the characteristic wavenumber and 
frequency of the wave-packet. The bounce time, $\tau_b$, for a particle in resonance with an
electromagnetic wave is proportional to its oscillation period in the pseudo-potential well governing the resonant  
wave-particle interaction \cite{KarimabadiEA90}. This interaction can be approximated by a Hamiltonian pendulum
in the vicinity of the resonance point.

Particles in resonance with a single finite amplitude fluctuation undergo a finite amplitude nonlinear oscillation.
This is the so-called trapping width, $\Delta v_{\parallel}$, given by the half peak-to-peak difference in the particle 
velocity parallel component. 
The trapping width and the bounce frequency for a nonrelativistic particle interacting resonantly with an 
electromagnetic wave are given by Equations (5a)--(5c) of Ref.~\cite{KarimabadiEA92}. 
These approximate expressions for $\Delta v_{\parallel}$ and $\omega_b$ 
yield considerable physical insight into the diffusion process \cite{MaceEA12} 
when used in conjunction with the quasilinear diffusion coefficient.

\section{Magnetic moment trapping width}
\label{sec:trap}

From the trapping width, $\Delta v_{\parallel}$, and bounce frequency, $\omega_b$, 
computed by Ref.~\cite{MaceEA12} for the case of a circularly polarized electromagnetic wave  
(see Appendix), it is possible to derive the pitch angle trapping half width as:
\begin{equation}\label{deltaalpha}
\Delta  \alpha =  \frac{\Delta  v_\parallel}{v} = 2 \left[(1-\alpha^2)^{1/2}|\alpha|\frac{\delta B}{B_0}\right]^{1/2} 
\end{equation}
As magnetic moment $\mu$ is related to $\alpha$ by Eq.~(\ref{mu}), we can write the trapping width 
for the magnetic moment as:
\begin{equation}
\label{mu_trap_a}
\Delta \mu = 2 \alpha \Delta \alpha = 4\alpha \left[(1-\alpha^2)^{1/2}|\alpha|\frac{\delta B}{B_0}\right]^{1/2}
\end{equation}
These expressions apply to a circularly polarized wave.
From Eq.~(\ref{mu_trap_a}) we expect that $\mu$ continues to be a good adiabatic invariant when resonances are not present or when particle interacts with extremely small amplitude waves.

\section{Model and governing equations}
\label{sec:model}

We investigate magnetic moment behavior first during the resonant interaction 
between one ion and a circularly polarized  magnetic wave, then when 
resonance overlapping occurs and finally during the interaction between 
a distribution of particles and a broad-band turbulent spectrum.
Because some of our normalization quantities are expressed in terms of typical time and length scales 
of the turbulence {\it slab} model~\cite{Jokipii66, BieberEA94}, we first give a general summary of the slab 
model.

For the general one dimensional (1D) slab description, turbulence is made up of a sum of right and left handed circularly polarized 
nondispersive plane  Alfv\'en waves propagating in the parallel direction. 
The magnetic field fluctuations are perpendicular to both the wave vector and the mean field. 
The fields are assumed to be magnetostatic. This amounts to the auxiliary assumption that the average particle 
speed is well in excess of the phase speed of the underlying linear wave mode. We ignore nonlinear wave-wave 
couplings in the spirit of quasilinear theory \cite[see e.g.,][]{KennelEA66, Swanson89, Stix92}. 

\begin {table}
\center
\begin{tabular}{lc}
\hline\hline\\[-8pt]
Arbitrary length scale                                &    $\lambda$ \\
Alfv\'en speed                                           &    $v_A$ \\
Unit transit time                                        &    $\tau_A = \lambda/v_A$ \\  
Magnetic field                                           &    $B_0 = \sqrt{4 \pi \rho}\ v_A$ \\
Electric field                                              &    $E_n = (v_A/c)B_0 = v_A^2 \sqrt{4 \pi \rho}/c$  \\[2pt]
\hline\hline
\end{tabular}
\caption{Characteristic physical quantities.}
\label{tab1}
\end{table}
Considering Alfv\'en waves propagating with $\omega/k = \omega/k_\parallel \simeq \pm v_A$, 
the magnetostatic approximation implies $|{\bf v}| \gg v_A$ (strictly $|v_\parallel | \gg v_A$). 
Since particle energy is conserved in a frame moving at the parallel component of the phase velocity of the wave
($\omega / k_\parallel$), quasilinear theory~\cite{KennelEA66} implies:
$$
(v_\parallel - \omega/ k_\parallel)^2 + {v_\perp}^2 = \mbox{const}.
$$
Because of the magnetostatic assumption, particle energy is conserved, i.e., 
energy diffusion in forbidden and in velocity space the resonant interaction diffuses pitch angle 
and gyrophase only. Finally, we ignore all inter-particle correlations resulting from their mutual interaction through their 
microfields (e.g.,Coulomb collisions, Debye shielding, and polarization). Furthermore 
the feedback of the particles on the macroscopic fields is ignored, i.e., we consider only test particles in prescribed 
macroscopic magnetostatic fields.
By virtue of the inequality $v_{\parallel} \gg v_A$, the turbulent electric field of the order $(\delta B/B_0) v_A B_0$ is negligible
compared to the motional electric field of the particle, $v_{\parallel}B_0$.

The dispersionless hypothesis rules out  phase mixing and, hence, phase decorrelation 
due to this process. Consequently the only way for a particle to see a ``wavepacket'' phase-decorrelate 
is to traverse an autocorrelation length of the turbulence \cite{KaiserEA78}. The autocorrelation time in this case is given by
\begin{equation}
\tau_{c} = \frac{1}{|\Delta(\omega - k_\parallel v_\parallel)|} =
\frac{1}{|v_\parallel \Delta k_\parallel|} \simeq
\frac{\lambda_c}{|v_\parallel|},
\end{equation}
where $\lambda_c$ is the turbulence correlation length.

The behavior of a test particle is described by its time dependent position ${\bf r}(t)$ and three-dimensional 
velocity ${\bf v}(t)$, that are advanced according to $d{\bf r}/dt = {\bf v}$ and the Lorentz force equation:
\begin{equation}
\label{eq:motion}
m\frac{d{\bf v}}{dt} = q\left[{\bf E} + \frac{\mathbf v}{c} \times {\bf B} \right]
\end{equation}

In order to render the equations non-dimensional, we use the characteristic quantities
listed in Table~\ref{tab1}, where
$\tau_A$ is the Alfv\'en crossing time, $v_A$ is the Alfv\'en velocity, $\lambda = l_z$ is the 
turbulence coherence length related to the turbulence correlation length $\lambda_c$ 
($\lambda_c = 0.747l_z$ for our particular slab configuration~\cite{MaceEA00}). 
For the static case also the light speed may be used as a  characteristic 
quantity~\cite{MinnieEA05}.
The introduction of an Alfv\'en speed in our test particle model, where the waves  are 
treated as static, may appear rather artificial. However, the magnetostatic assumption is valid here 
provided that $|v_\parallel | \gg v_A$ and we introduce $v_A$ in anticipation of future work 
where we will drop the magnetostatic hypothesis.

With our choice for the characteristic quantities (Table~\ref{tab1}) the dimensionless 
equations of motion of our charged test particles are given by:
\begin{eqnarray} \label{ODE}
\frac{d{\bf r}}{dt}            &     =     & {\bf v}\\
\frac{d{\bf v}}{dt}           &     =     & \beta({\bf E} + {\bf v} \times {\bf B})
\end{eqnarray}
Here $\beta=\Omega \tau_A$ \cite[cf.~$\alpha$ parameter in Ref.][]{AmbrosianoEA88} couples particle and 
field spatial and temporal scales and provides a particularly useful means to relate our numerical 
experiments to space and astrophysical plasmas. 
In general in a turbulent collisionless plasma the bandwidth of the inertial range fluctuations 
may extend from large fluctuations at the correlation scale, $\lambda_c$, to small fluctuations 
at the ion inertial scale. In this case $\beta \gg 1 $ and
the turbulent time-scales are much slower than the typical particle gyroradius~\cite{GoldsteinEA86}.

The resonant condition for the static case in terms of $\beta$ is given by
\begin{equation}\label{res_cond}
k_{res}\lambda= \frac{n\beta}{\alpha (v/v_A)} =  \frac{n\beta}{(v_\parallel/v_A)} 
\end{equation}

Time is advanced through a fourth-order Runge-Kutta integration method with an adaptive 
time-step~\cite[pp. 708-716 of Ref.][]{PressEA92}.

\section{Numerical simulations}
\label{sec:simulations}

Particles are loaded randomly in space at $t = 0$ throughout a one-dimensional simulation 
box of length $L$. The fields are described in the following sections. 
In spherical coordinates, with the polar axis along the $z$-direction parallel to the mean 
magnetic field of strength $B_0$, particle velocity components are:
\begin{equation}
v_x = v\sin \theta \cos \phi \quad v_y = v\sin \theta \sin \phi  \quad v_z = v\cos \theta 
\label{velocities}
\end{equation}
Particles initial velocities are randomly distributed in the gyrophase $\phi$ between $[0:2\pi]$, while
the velocity magnitude $v$ and pitch angle $\theta$ are determined by the particular numerical experiment.

Typical particle velocities used in our simulations are $10v_A$ and $100v_A$, satisfying the magnetostatic constraint. 
In our analysis magnetic moments are expressed in units of the characteristic quantity
$\mu_n= v^2/B_0$. We also define $\delta b = \delta B/B_0$.

\begin{figure}
\begin{center}
\includegraphics[width=8.6cm]{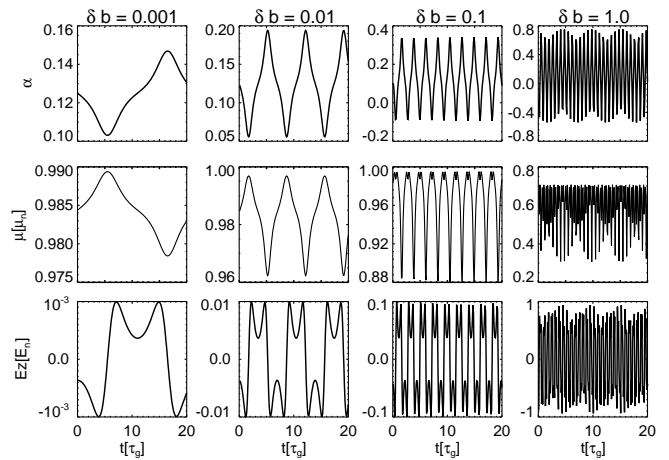}
\end{center}
\caption{Gyroresonant interaction between a circularly polarized wave 
and a particle with $v=100 v_A$ and $\alpha=1/8$: cosine of pitch angle $\alpha$ (top row), particle magnetic 
moment $\mu$ (middle row) and parallel component of the induced electric field bottom (row). 
Different columns correspond to different wave amplitude: 
$\delta b = 0.001$ (first column), $\delta b = 0.01$ (second column), $\delta b = 0.1$ (third 
column) and $\delta b = 1.0$ (fourth column).}
\label{ame_trap}
\end{figure}

The statistic analysis of particle magnetic moment involves averaging trajectories over the particle gyroperiod
$\tau_g = 2\pi / \Omega$. For each simulation we compute the effective number of gyroperiods $N_{\tau_g}$
that particles complete in a given magnetic field configuration as:
\begin{equation}
N_{\tau_g} = \int_0^t \frac{dt}{2\pi} \frac{eB(t)}{mc}.
\label{eq:totaltau}
\end{equation}
where $B(t)$ is the intensity of the total magnetic field. When $\delta b \ll 1$, $B(t) \simeq B_0$;
however increasing $\delta b$ toward unity the waves contribution to the 
strength of the total magnetic field $B(t)$ is not negligible.

\subsection{Single wave}
\label{sec:onewave}

We start studying the ion motion in presence of a constant magnetic field ${\bf B}_0$ and a 
perpendicular  left-handed circularly polarized wave with
\begin{equation} \label{Bfield}
{\bf B} = \delta B_x \cos(k_0z)\, \mathbf{\hat{e}}_x - \delta  B_y \sin{(k_0z)}\, \mathbf{\hat{e}}_y + 
B_0\, \mathbf{\hat{e}}_z,
\end{equation}
where $\delta B_x$ and $\delta B_y$ are the amplitudes of the wave and $k_0$ is the wavevector. 
We assume $\delta B_x = \delta B_y = \delta B$ for the rms average values.
In these simulations $\beta=10^3$, $v = 100v_A$ and $\alpha = 0.125$ ($\theta = 82^\circ$). 

We follow the test-particles until they complete $N_{\tau_g} = 100$ gyroperiods. 
For the resonance condition, Eq.~(\ref{res_cond}), we set $k_0=80 /\lambda$.
Particles injected with a pitch angle cosine different to $\alpha = 0.125$ will not be in resonance with this wave, exhibiting
a different behavior. For a direct comparison we also inject non-resonant particles, i.e., with 
$\alpha = 0.5$ ($\theta = 60^{\circ}$).

Figure~\ref{ame_trap} shows the time evolution of the cosine of pitch angle $\alpha$,  
particle magnetic moment $\mu$, and the parallel component of the induced electric 
field $E_z$, for a resonant particle ($\alpha = 0.125$). 
Different columns corresponds to different values of the wave amplitude: $\delta b = 0.001$ (first column), 
$\delta b = 0.01$ (second column), $\delta b = 0.1$ (third column) and $\delta b = 1.0$ (fourth column). 

When the parallel component of the induced electric field is almost constant and equal to 
$E_z \sim -v_{\perp}\delta b$, the resonant interaction produces variations that are secular over a gyroperiod.
However an oscillation occurs over a longer time, the bounce period 
$\tau_b=2\pi/ \omega_b$ (where $\omega_b$ is the bounce frequency discussed in Section~\ref{sec:trapp_theory}). 
This is the typical timescale over which the velocity, and hence the particle trajectory, 
exhibits significant deviations from the linear $v_{\parallel} = \text{const}$ and $v_{\perp} = \text{const}$ case.

\begin{table}
\center
\begin{tabular}{ | l | c | c | c | c |}
\hline
$\delta b$  &  $\Delta\alpha_{th}$   &    $\Delta\alpha_{sim}$       &     $\Delta\mu_{th}$    &    $\Delta\mu_{sim}$       \\            
\hline
$0.001$              &    $0.022$        &       $0.02$                             &           $0.0056$         &      $0.0055$ \\ 
$0.01$                &    $0.07$          &       $0.075$                           &           $0.0176$         &      $0.02$ \\ 
$0.1$                  &    $0.2227$      &       $0.2$                               &           $0.0556$         &      $0.055$ \\ 
$1.0$                  &    $0.704$        &       $0.6$                               &            $0.176$          &      $0.175$ \\ 
\hline
\end{tabular}
\caption{Trapping width values for $\alpha$ and $\mu$:
comparison between theoretical (subscript \emph{th}) and numerical (subscript \emph{sim}) values.}
\label{tab_trap}
\end{table}
In Section~\ref{sec:trap} we derived the analytical expression for the half trapping-width 
of magnetic moment for a particle interacting with a left or right handed circularly polarized wave (see Eq.~\ref{mu_trap_a}). 
We now compute the values of the half peak-to-peak difference in $\alpha$ and $\mu$, 
$\Delta \alpha = (\alpha_{max} - \alpha_{min})/2$ and $\Delta \mu = (\mu_{max} - \mu_{min})/2$, for 
the resonant interaction simulations. These values and those obtained from the theoretical 
expressions~(\ref{deltaalpha})-(\ref{mu_trap_a}) are listed in Table~\ref{tab_trap} and
are in good agreement, confirming the validity of equations~(\ref{deltaalpha})-(\ref{mu_trap_a})
and reinforcing the intuitively idea that magnetic moment and pitch angle behaviors are strictly related.

\begin{figure}
\begin{center}
\includegraphics[width=8.6cm]{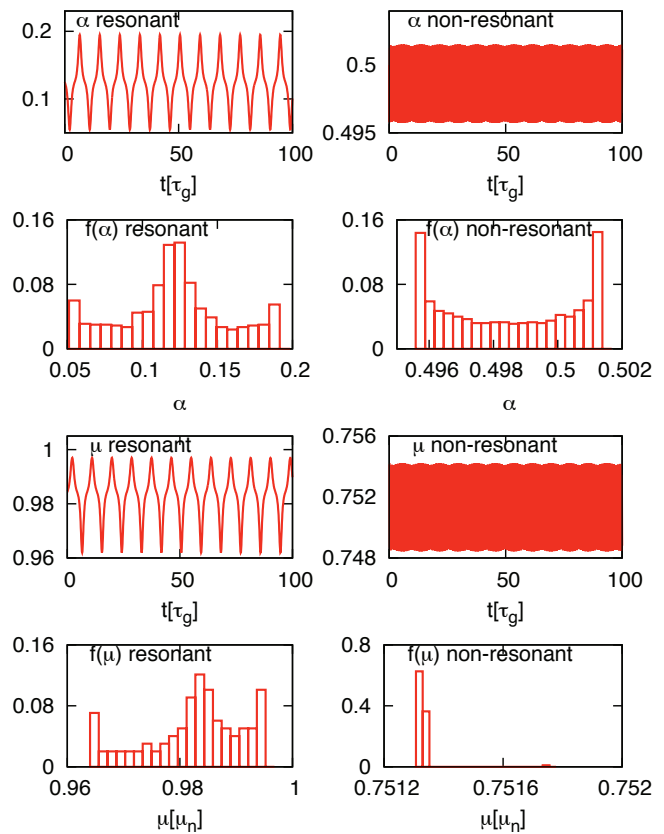}
\end{center}
\caption{Time evolution of cosine of pitch angle  $\alpha$ (first row) and its distribution function  $f(\alpha)$ (second row),
time evolution of magnetic moment $\mu$ (third row) and its distribution function $f(\mu)$ (fourth row) of \emph{resonant}
($\alpha = 0.125$, left column) and \emph{non-resonant} particle ($\alpha = 0.5$, right column). $v=100\, v_A$.}
\label{res_vs_nres}
\end{figure}

To compare resonant and non-resonant dynamics, we show in Figure~\ref{res_vs_nres} the time evolution of cosine 
of pitch angle $\alpha$ (first row), magnetic moment $\mu$ (third row), and their distribution functions 
$f(\alpha)$ (second row) and $f(\mu)$ (fourth row) at the end of the simulation, 
for a \emph{resonant particle} with $\alpha = 0.125$ 
(left column), and a \emph{non-resonant} one with $\alpha = 0.5$ (right column). 
In contrast with the resonant case in which $\alpha$  and $\mu$  exhibit  well-known secular 
variations with typical period equal to $\tau_b$, the $\alpha$ and $\mu$ profiles for a non-resonant 
particle show a regular oscillating behavior, a distinctive signature of regular particle motion. 
The values of the half peak-to-peak difference in $\alpha$ and $\mu$ obtained from the simulation are $\Delta \alpha_{sim} =0.0025$ and 
$\Delta \mu_{sim} = 0.003$. These are smaller than the theoretical values computed from
equations~(\ref{deltaalpha})-(\ref{mu_trap_a}) with $\delta b=0.01$ and $\alpha = 0.5$, for which we obtain 
$\Delta \alpha_{th} =0.1316$ and $\Delta \mu_{th} = 0.1316$. 

The distribution functions $f(\alpha)$ and $f(\mu)$ (Figure~\ref{res_vs_nres}) for a resonant 
particle are more spread in $\alpha$ and $\mu$ and are centered around their initial values 
$\alpha = 0.125$ and $\mu = 0.98$. In the non-resonant case, $f(\mu)$ remains peaked at its 
initial value, i.e., its magnetic moment is constant during particle motion. 
The spread in $\alpha$ of its distribution is $\sim10\%$, small compared to the resonant 
case spreading of $\sim40\%$.

\begin{figure}
\begin{center}
\includegraphics[width=8.6cm]{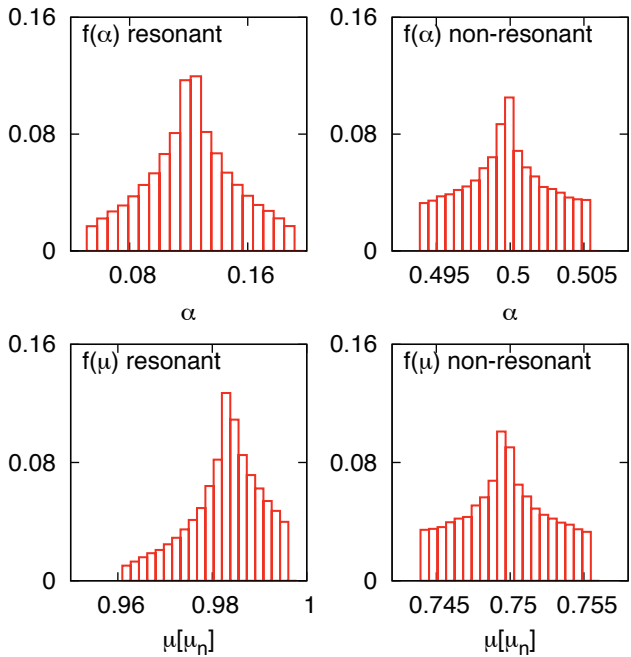}
\end{center}
\caption{$f(\alpha)$ (first row) and $f(\mu)$ (second row) at the end of the simulation
for an initial distribution of $1000$ \emph{resonant} (left column) and \emph{non-resonant} (right column) particles 
randomly distributed in the simulation box. $\delta b=0.01$.}
\label{res_vs_nres_1000}
\end{figure}

Figure~\ref{res_vs_nres_1000} shows the distribution functions, $f(\alpha)$ and $f(\mu)$,
at the end of the simulation for $1000$ resonant and non-resonant particles
injected in the simulation box with random positions and  
phases. For non-resonant particles (right column) the distributions remain peaked 
around their initial values $\alpha = 0.5$ and $\mu/\mu_n = 0.75$ with very little spreading. 
For the resonant particles (left column) $f(\alpha)$ acquires a Gaussian shape centered around its initial 
value $\alpha = 0.125$. Furthermore it spreads of $\sim 0.1$, comparable to the trapping width 
for the single particle $2\Delta \alpha = 0.014$ (Figure~\ref{res_vs_nres}).  
The magnetic moment distribution for the resonant case has a characteristic shape
found for $\mu$ in the parameter range in which pitch angle exhibits a 
Gaussian distribution and the density distribution function is still isotropic 
(particle free-streaming regime).
As for the pitch angle, the spread in the magnetic moment distribution of $\sim 0.03$ is comparable 
to the trapping width for the single particle $2\Delta \mu = 0.00352$ (see Eq.~\ref{mu_trap_a} 
and Figure~\ref{res_vs_nres}).

\subsection{Overlapping resonances}
\label{sec:overlap}

In order to understand the effect of overlapping resonances on particle magnetic moment,
we perform a numerical experiment 
with four different particles in the simulation box with random initial positions, 
same initial velocity $v = 100\, v_A$, but different values for pitch angle cosine: 
$\alpha_1=1/2, \alpha_2=1/4,\alpha_3=1/8, \alpha_4=1/32$. 
For $\beta=10^3$, making use of the resonance condition for the static case [Eq.~(\ref{res_cond})],
the cyclotron resonances $n=1$ for the different values of $\alpha$  are expected for
$k_1 \lambda=20$, $k_2\lambda=40$, $k_3\lambda=80$, and $k_4\lambda=320$. 

The total magnetic field is given by:
\begin{equation}
{\bf B} =B_0{\bf \hat{e}}_z + \sum_{i=1}^{4} \delta b \cos[k_iz + \phi_i] {\bf \hat{e}}_x - \sum_{i=1}^{4} 
\delta b \sin[k_iz + \phi_i] {\bf \hat{e}}_y,
\end{equation}
where the $\phi_i$ are random phases.
Taking into account resonance broadening effects, all particles with parallel velocities in the range 
\begin{equation}\label{range}
v_{\parallel} - \Delta v_{\parallel} < v_{\parallel} < v_{\parallel} + \Delta v_{\parallel}
\end{equation}
can potentially resonate with a wave, whose wave number is $k_{\parallel}=\Omega/v_{\parallel}$. 
As found by Ref.~\cite{Chirikov78}, the direct evidence of resonances overlapping is the disappearance 
of constants of motion, i.e., the onset of stochasticity in the Hamiltonian formalism. 
We make simulations with four different waves amplitudes $\delta b$ = 0.001, 0.01, 0.1, and 1.0.  
The values of the trapping half-widths $\Delta v_{\parallel}$ computed for the different pitch angles 
with Eq.~(\ref{deltav}) are listed in Table~\ref{tab_overlap} for the different $\delta b$ considered.
\begin{table}
\center
\begin{tabular}{ | l | c | c | c | c |}
\hline
$\delta b$     &    $\alpha=1/2$      &    $\alpha=1/4$      &  $\alpha=1/8$  &  $\alpha=1/32$  \\            
\hline
$0.001$        &    $4.16$                &       $3.1$               &    $2.227$        &      $1.3$        \\ 
$0.01$          &    $13.1$                &       $9.85$             &   $7.042$         &      $3.583$    \\ 
$0.1$            &    $41$                   &       $31.1$             &    $22.27$        &      $11.33$     \\ 
$1.0$            &    $131$                 &       $98.3$             &    $70.42$        &      $35.83$     \\ 
\hline
\end{tabular}
\caption{Values of $\Delta v_{\parallel}$ 
for $\alpha$ =1/2,\, 1/4,\, 1/8,\, and 1/32 resonances at different $\delta b$.}
\label{tab_overlap}
\end{table}

\begin{figure}
\begin{center}
\includegraphics[width=8.6cm]{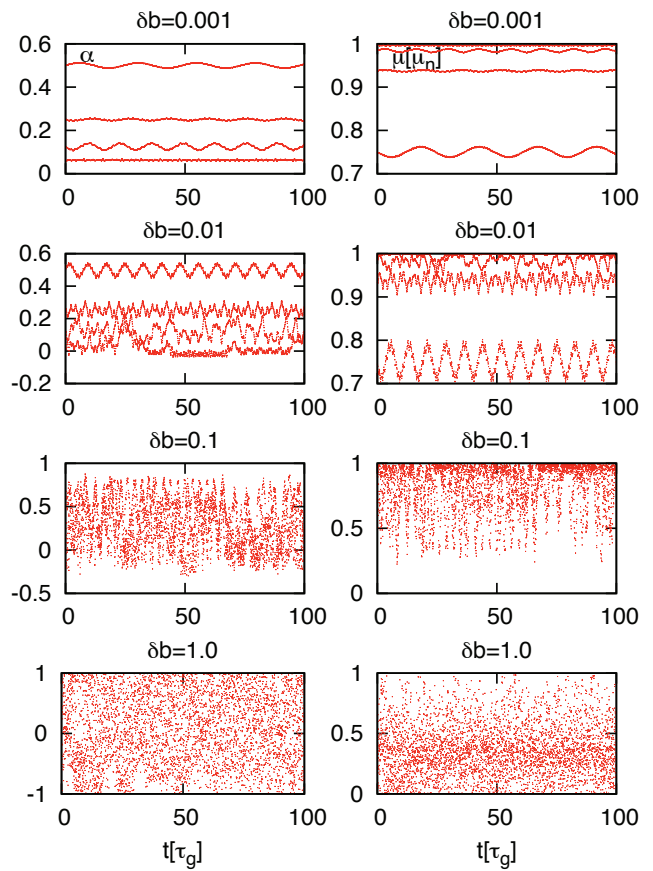}
\end{center}
\caption{Transition from non-overlapping to overlapping resonances: $\alpha$ (left column) and 
$\mu$ (right column) profiles varying the waves amplitude: $\delta b=0.001$ (first row), $\delta b=0.01$ 
(second row), $\delta b=0.1$ (third row), $\delta b=1.0$ (fourth row).}
\label{overlap}
\end{figure}
Figure~\ref{overlap} shows time histories of pitch angle cosine $\alpha$ (left column)
and magnetic moment $\mu$ (right column) profiles for various $\delta b$.
Again similar behavior is seen for $\alpha$ and $\mu$. 
For the smallest wave amplitude, $\delta b=0.001$, it is possible to recognize 
very well the four different resonances in the profiles of $\alpha$ and $\mu$. 
For $\delta b=0.01$ the resonance at $\alpha_3=1/8$ is overlapping with the 
resonance at $\alpha_4=1/32$. Indeed, the initial parallel velocity of the 
particle injected at the smallest pitch angle, $v_{\parallel,4}=3.125v_A$, lies 
in the range of velocities [see Eq.~(\ref{range})] in possible resonance with 
$k_{\parallel}=k_3$. For higher wave amplitudes, $\delta b=0.1$ 
and $\delta b=1.0$, the condition~(\ref{range}) is satisfied by all particles velocities. 
Stochasticity arises and the different resonances are indistinguishable.

\begin{figure}
\begin{center}
\includegraphics[width=8.6cm]{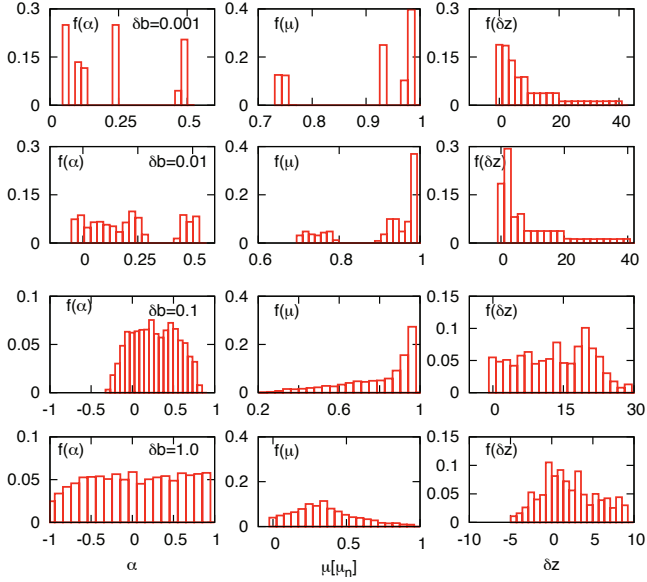}
\end{center}
\caption{Transition from non-overlapping to overlapping resonances with varying wave amplitude: $\delta b=0.001$ 
(first row), $\delta b=0.01$ (second row), $\delta b=0.1$ (third row), 
$\delta b=1.0$ (fourth row). The distribution functions $f(\alpha)$ (left column), $f(\mu)$ (central column) and 
$f(\delta z)$ (right column) are averaged over time.}
\label{overlap_pdf}
\end{figure}

The distribution functions $f(\alpha)$, $f(\mu)$ and $f(\delta z)$ 
(where $\delta z=z - z_0$ is the displacement along $z$ relative to the
particle initial position $z_0$) after $100 \tau_g$ (Figure~\ref{overlap_pdf}) exhibit 
similar characteristics.

For $\delta b = 0.001$, $f(\alpha)$ and $f(\mu)$ are peaked 
in correspondence of their four initial values because of the good 
resonances separation. $f(\delta z)$ shows that the particles are 
simply free-streaming in the parallel direction and, depending on 
their initial parallel velocity, they cover shorter or longer distances along $z$. 

For $\delta b=0.01$, $f(\alpha)$ spreads around its initial four peaks 
because particle interact resonantly with waves of larger amplitude,
and resonances overlap for $\alpha < 1/4$, as discussed previously. 
Similar effects are shown also by $f(\mu)$, confirming that for small 
$\delta b$ the resonant interaction affects magnetic moment and 
pitch angle in similar ways. 

While for $\delta b=0.01$ particles continue to free-stream in the z-direction,
different profiles for  $f(\delta z)$ appear for $\delta b=0.1$. 
Pitch angle distribution begins to isotropize and magnetic moment 
exhibits a one-sided long tail distribution extending toward smaller $\mu$. 
This behavior is similar to the regime found previously in the single wave 
experiment when $f(\alpha)$ is nearly isotropic, $f(\delta z)$ still 
indicates particles free-streaming, and the magnetic moment distribution 
displays a long tail. 

For $\delta b=1.0$, by $t = 100 \tau_g$ the
pitch angle cosine distribution  $f(\alpha)$ has become completely isotropic, while 
$f(\delta z)$ approaches a gaussian distribution indicative of spatial diffusion.
In this regime $f(\mu)$ loses its long-tail and starts to acquire a gaussian shape. 
In that way we have identified three distinct regimes of statistical magnetic 
moment behavior with increasing degree of turbulence.

\section{Slab spectrum}

\begin{figure}
\begin{center}
\includegraphics[width=8.5cm]{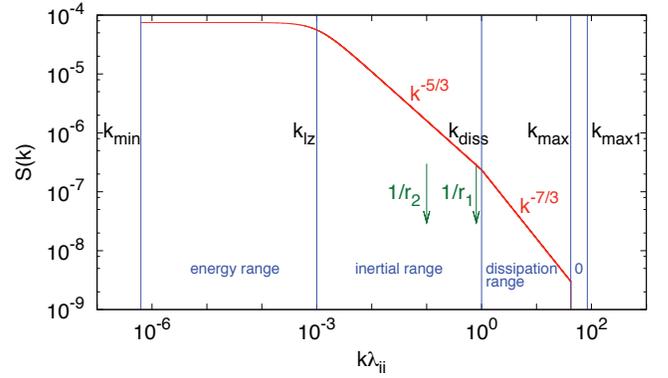}
\caption{Power spectrum of the turbulent magnetic. $k$ is normalized to the coherence length $l_z$.}
\label{fig:sk}
\end{center}
\end{figure}

In this section we present the results of our numerical simulations of test-particles in presence of a 
broad-band slab spectrum [see Eq.~(\ref{eq:pk}) and Figure~\ref{fig:sk}]. 
We have performed simulations for different 
particles velocities and amplitude of the magnetic field fluctuations.

Simulations use a unidimensional computational box of length $L = 10000\, l_z$ ($l_z = 1$ is the coherence scale for the slab spectrum) with
$N_{z} = 2^{28} = 268,435,456$ grid points. 
The magnetic field in physical space is generated from a spectrum $P(k)$ in Fourier space,
via inverse fast Fourier transform (FFT). 
The turbulent magnetic field is given by:
\begin{equation}
{\bf B}(z) = B_0{\bf e}_z + \delta {\bf B}(z),
\label{eq:totalb}
\end{equation}
with $\delta {\bf B}(z) = \delta B_x(z)\, {\bf \hat{e}}_x + \delta B_y(z)\, {\bf \hat{e}}_y$
and the solenoidality condition is identically satisfied.

The modes of the magnetic field components in k-space are given by:
\begin{align*}
\delta B_x(k_n)= [P(k_n)]^{1/2} e^{i\Phi_n} \\
\delta B_y(k_n) = [P(k_n)]^{1/2} e^{i\Psi_n}
\end {align*}
where $k_n = 2\pi n/L$ and $\Phi_n$ and $\Psi_n$ are random  phases.
The slab spectrum $P(k)$ is given by:
\begin{equation}
P(k_n) =
\begin{cases}
 C_{slab} [1 + (k_n l_z)^2]^{-5/6}, & \text{for $k_n < k_{diss}$} \\
C_{diss} \left( \frac{k_n}{k_{diss}} \right)^{-7/3}, & \text{for $k_n \ge k_{diss}$}
\end{cases}
\label{eq:pk}
\end{equation}
where $C_{slab} = 2\lambda_c \delta b^2_{x,slab}$ is  a constant specific to this form of the slab model, 
$\delta b^2_{x,slab}$ is the mean square fluctuation, $k_{diss}$ is the dissipation range wavenumber,
$C_{diss} = C_{slab}[1 + (k_{diss} l_z)^2]^{-5/6}$ is the constant for the dissipation range
(set by the continuity of the spectrum $P(k)$ at $k_{diss}$).
The vectors of Fourier coefficients are zero-padded for $N_{max} + 1 \le n \le N_z$
providing an extra level of smoothness to the fields by an effective trigonometric interpolation. 
In all the simulations we use $N_{max} = 6.7 \times 10^7$ and a simple linear interpolation 
to compute the fields at the test particle position. 

The resulting spectrum is shown in Figure~\ref{fig:sk}.
Several important scales are present in the system. They are labeled as 
$k_{min}$, $k_{l_z}$, $k_{diss}$, $k_{max}$ and $k_{N_z}$. The discrete wavenumbers are obtained
through $k_n = 2 \pi n/ L$ as:
\begin{equation}
N_k  = \frac{L}{2 \pi} k \sim 1600\, k.
\label{eq:nk}
\end{equation}

\begin {table}
\caption{Characteristic scales in the spectrum.}
\center
\begin{tabular}{lcr}
\hline
Wavenumber index                                &    Wavenumber value \\
\hline
$N_{kmin} = 1$                                                    &    $k_{min} = 6.28 \times 10^{-4}$ \\
$N_{kl_z} = 10^4$                                                &    $k_{l_z} = 2\pi$ \\  
$N_{k_{diss}} = 1.6 \times 10^6$                         &    $k_{diss} = 10^3$ \\ 
$N_{k_{MAX}} = 6.7 \times 10^7$                       &    $k_{MAX} = 4.2 \times 10^4$ \\ 
$N_{k_{MAX1}} = 1.3 \times 10^8$                     &    $k_{MAX1} = 8.4 \times 10^4$ \\ 
\hline
\end{tabular}
\label{tab:nk}
\end{table}

We summarize the values for $k$ and $N_k$ used in our simulations in Table~\ref{tab:nk}, where:
\begin{itemize}
\item[-] $k_{min} = 2\pi /L$ is the minimum wave vector of the spectrum, corresponding to $N_k = N_{kmin} = 1$. 
\item[-] $k_{l_z} = 2\pi/l_z =2\pi$ is the wave vector that marks the beginning of the \emph{inertial range}. 
Three decades of \emph{energy containing range} from $k_{min}$ to $k_{l_z}$
ensure turbulence homogeneity. $l_z$ or $\lambda_c = 0.747l_z$ \cite[see Ref.][]{MaceEA00} correspond to
the typical lengths scales over which the particles attain diffusive behavior of the pitch angle.
Three decades of inertial range with $P(k) \propto k^{-5/3}$ well represent solar wind conditions. 
\item[-] $k_{diss}$ is the wave vector corresponding to the beginning of the \emph{dissipation range}. 
In our model, the spectrum extends beyond $k_{diss}$ with $P(k) \propto k^{-7/3}$. 
\item[-] At two decades higher wavenumber, $k_{MAX} = \sqrt{m_i/m_e} k_{diss}$ determines the end of the dissipation range. 
\item[-] Extending for two decades beyond $k_{MAX} = 4.2 \times 10^4$,
the spectrum includes zero-padding up to $k_{MAX1} = 8.4 \times 10^4$. 
\item[-] Another important scale, not labeled in Figure~\ref{fig:sk} because it depends on test-particle velocity, 
is the wave vector corresponding to $z_{max} = vT_{tot}$, the distance covered by a charged test particle 
moving at speed $v$ in the simulation running time $T_{tot}$. 
To avoid periodicity effects it is important that the box length $L$ is large enough so that particles trajectories 
are limited to a small fraction of the full length, i.e., $L \gg z_{max}$ or $k_{min} \ll 1/z_{max}$. 
Periodicity might indeed give rise to artificial field lines diffusion.
\end{itemize}

We fix the value of the $\beta$ parameter equal to $10^4$.
This corresponds approximately to observed solar wind turbulence properties at $1$~AU, as follows:
\begin{equation}
\beta =  \Omega \tau_A = \left(\frac{q}{m}\right) \frac{\lambda_c \sqrt{4\pi \rho}}{c} = \frac{\omega_{pi}\lambda_c}{c} = \frac{\lambda_c}{\lambda_{ii}},
\end{equation}
where $\lambda_c$ is the turbulence correlation length 
$\omega_{pi}= (4\pi n_{0i}{q_i}^2/m_i)^{1/2}$ is
the ion plasma frequency ($q_i$ and $m_i$ are respectively the ion charge and mass)
and $\lambda_{ii} = c/\omega_{pi}=\left( c^2 m_i / 4 \pi n_i e^2 \right)^{1/2}$ is the ion inertial length. 
For $n_i = n_e$ then $\lambda_{ii} = (m_i/m_e)^{1/2}\rho_{ie}$. 
Because the solar wind density at 1~AU\ is approximately $n \sim (1,10)\, cm^{-3}$
on average $\lambda_{ii} \sim 1000\, km$. 
At the same distance the turbulence correlation length $\lambda_c$ 
is approximately $10^6$\,km \cite{MatthaeusEA86} and $\beta \simeq 10^4$.

Typically $1000$ particles are injected in the simulation with initial random positions. 
Particles are loaded from a cold ring beam [see equations~(\ref{velocities})] distribution with constant velocity magnitude, 
$\sin \theta$ is set equal to $(1 - \alpha_0^2)^{1/2}$, where $\alpha_0$ is the initial pitch angle 
cosine respect to the background field $B_0$. The initial gyrophase $\phi$ is chosen randomly. 
For all the simulations $\alpha_0=0.125$ ($\theta \simeq 82^{\circ}$). 

\begin{table}
\caption{Typical values used in the simulations.}
\center
\begin{tabular}{ | l | c | c | c | c |}
\hline
$V[v_A]$                        &    $r_L[l_z]$                                        &    $k_{res}$                              &                            $\epsilon$                                       &    $t_c[\tau_A]$  \\            
\hline
$10$                                  &    $10^{-3}$                                         &       $8\times10^{3}$                  &                          $1.33\times10^{-3}$                         &      $0.0747$      \\ 
$100$                               &    $10^{-2}$                                         &       $8\times10^{2}$                  &                           $1.33\times10^{-2}$                        &      $0.00747$        \\      
\hline
\end{tabular}
\label{tab:part}
\end{table}

From the previous section, we know that the behavior of magnetic moment is correlated to pitch angle behavior for a low level of magnetic fluctuation ($\delta b =0.001, 0.01$). 
Pitch angle and magnetic moment exhibit Gaussian distribution functions typical of 
normal diffusion processes. Increasing the turbulence level, pitch angle distribution approaches 
isotropization and a transient regime is observed with the magnetic moment starting to be 
influenced by the onset of spatial parallel diffusion. 
When $f(\alpha)$ completely isotropizes, spatial diffusion sets in and $f(\mu)$ behavior is closely 
related to the sampling of the varying magnetic field strength associated with that spatial diffusion. 

From quasilinear theory we know that velocity and real space diffusion occur at two different time scales. 
Typically, velocity space diffusion takes place with the time scale $\tau_c = \lambda_c/v$ 
shorter than the typical time scale at which parallel diffusion occurs $\tau_{\parallel} = \lambda_{\parallel}/v$,
where $\lambda_{\parallel} = 3D_\parallel/v$ is the parallel mean free path. 
For this reason we follow test particles in the simulation box for a time $T > \tau_c$,
typically with $T = 20\tau_c$. Particles parameters used in the simulations are listed in Table~\ref{tab:part}. 

An important parameter in the description of energetic test particles is $\epsilon = r_L/\lambda_c$, 
which is sometimes called the dimensionless particle rigidity. It can be related to the bend-over wavenumber of the turbulence,
$k_{bo} = 1/\lambda_c$, and the minimum resonant wavenumber, $k^{r}_{min} = 1/r_L$, 
as $\epsilon = k_{bo}/k^{r}_{min}$.
For example when $r_L \gg \lambda_c$ particles experience all 
possible $k$-modes in few gyroperiods resonating with the energy containing scale ($k^{r}_{min} \ll k_{bo}$). 
For lower energies the test particles resonate in the inertial range.
Those with  $v = 10\, v_A$ will resonate at the end of the inertial range ($1/r_1$ in Fig.~\ref{fig:sk}),
while those with $v\, = 100v_A$ at the middle of the inertial range ($1/r_2$ in Fig.~\ref{fig:sk}). 
Furthermore, as explained previously, the condition $k_{min} \ll z_{max}$ is necessary to avoid artificial effects
in particle transport associated with periodicity of the magnetic field. 
\begin{figure*}
\begin{center}
\includegraphics[width=16cm]{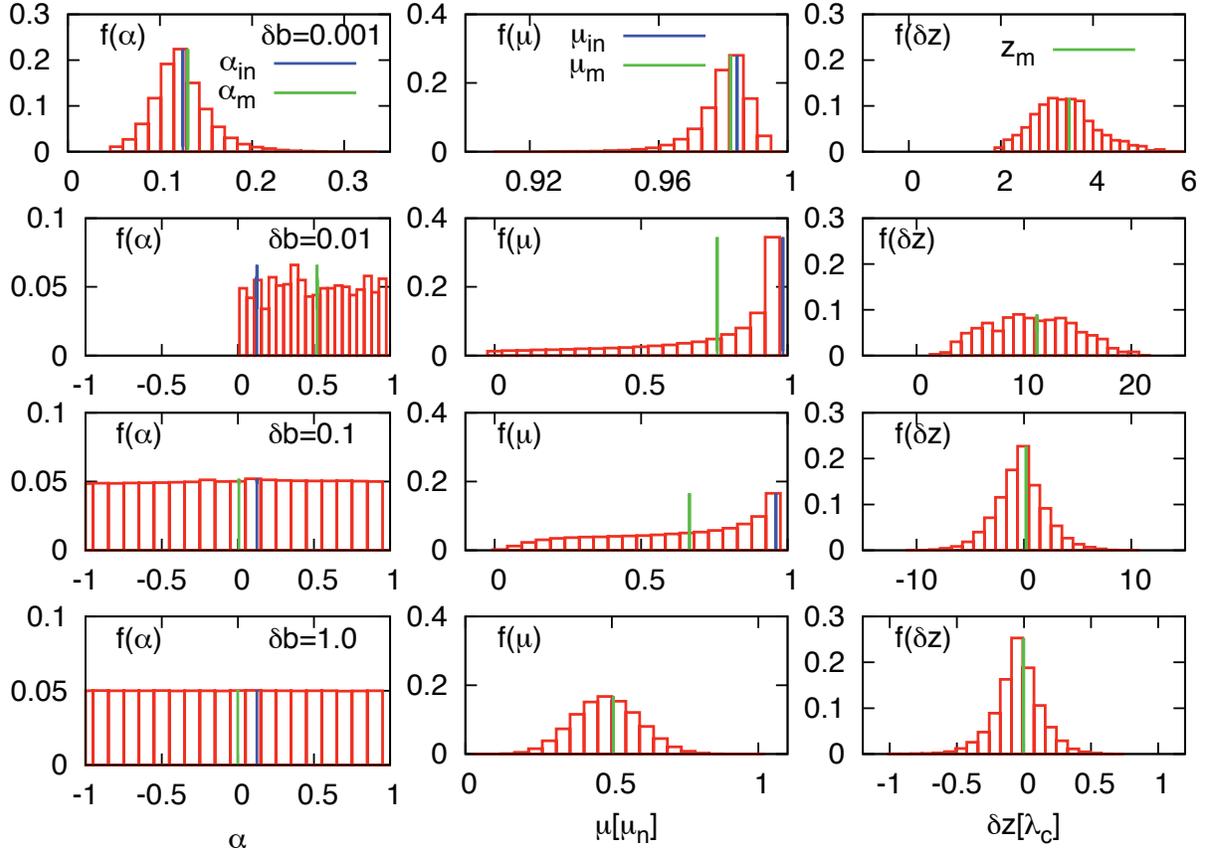}
\caption{Distribution functions of cosine of pitch angle $f(\alpha)$ (left column), magnetic moment $f(\mu)$ (central column) and particle displacements relative to the initial position $f(\delta z)$ (right column) at $20\, \tau_c$
for different waves amplitude: $\delta b=0.001$ (first row), $\delta b=0.01$ (second row), $\delta b=0.1$ (third row) 
and $\delta b=1.0$ (fourth row). Particle parameters at injection: $v = 10\, v_A$ and $\alpha_0=0.125$.}
\label{fig_10vadistr}
\end{center}
\end{figure*}

\begin{figure*}
\includegraphics[width=16cm]{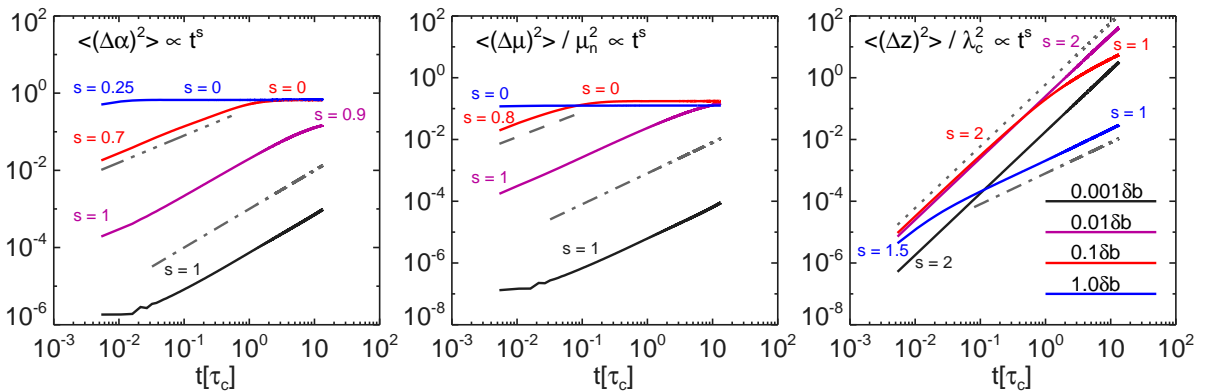}
\caption{Statistics for $v = 10\, v_A$. Variances for cosine of pitch angle $\alpha$ (left plot), magnetic moment $\mu$ (center plot) and particle displacements relative to the initial position $\delta z$ (right plot) 
at different values of $\delta b$: $\delta b=0.001$ (black line), $\delta b= 0.01$ (purple line), 
$\delta b = 0.1$ 
(red line) and $\delta b = 1.0$ (blue line). The $s$ values indicate the different slopes for the 
scalings $\langle( \Delta \alpha )^2 \rangle, \langle( \Delta \mu )^2 \rangle$ and
$\langle( \Delta z )^2 \rangle \propto t^s$. The variances are fitted with the gray lines:
dotted line is $s=2$, dotted-dashed line $s=1$, dashed line $s=0.8$, three dotted-dashed line $s=0.7$.}
\label{fig:10va}
\end{figure*}

Figure~\ref{fig_10vadistr} shows $f(\alpha)$ (left column), $f(\mu)$ (central column) and $f(\delta z)$ 
(right column) for a distribution of particles moving with an initial velocity $v = 10\, v_A$ in presence 
of the slab spectrum [Eq.~(\ref{eq:pk}), Figure~\ref{fig:sk}]. 
All the distribution functions are computed at the end of the simulation, i.e., after $20\, \tau_c$.
The blue line and the green line indicate the initial value and the mean value of each distribution.
As particles are injected at different positions, it is convenient to define the quantity $\delta z = z(j) - z(0)$
($j$ is a temporal index). In this way it is possible to take out from the distribution function $f(\delta z)$ 
both the drift effect ($v_Dt$) and particle diffusion relative to their own positions ($\Delta z_i$). The 
general expression for the $z$ position of the $i$-th particle is given by
\begin{equation}
z_i = z_i(0) + v_Dt + \Delta z_i = z_i(0) + \delta z_i.
\end{equation}

The primary diagnostic for studying particle diffusion is the variance $\sigma^2(t) \propto t^s$ of particles cosine of pitch angle, magnetic moment and position parallel to the mean field direction.
Figure~\ref{fig:10va} illustrates the time evolution of the variances, $\langle \left( \Delta \alpha \right)^2 \rangle$ (left figure), $\langle \left( \Delta \mu \right)^2 \rangle$ 
(central figure) and $\langle \left( \Delta z \right)^2 \rangle$ (left figure) for a particles distribution moving with initial velocity equal to $10\, v_A$.
Different colors correspond to different $\delta b$ values:  $\delta b=0.001$ black line, $\delta b= 0.01$ purple line, $\delta b = 0.1$ 
red line and $\delta b = 1.0$ blue line. The variances are fitted with the gray lines: the dotted line is used for $s=2$, the dotted-dashed line for $s=1$, the dashed line for $s=0.8$ and the three dotted-dashed line for $s=0.7$.

For $\delta b=0.001$, $\alpha$ and $\mu$ display Gaussian distributions while particles free-stream in the $z$-direction.
Particles that cover greater distance in $z$ are more scattered in pitch angle and consequently in $\mu$. 
Figure~\ref{fig:10va} shows superdiffusive behavior (black line, $s = 2$) with particles free streaming along $z$, and later, variance characteristic of normal diffusion with $\langle \left( \Delta \alpha \right)^2 \rangle$ and $\langle \left( \Delta \mu \right)^2 \rangle$ scaling $\propto t$.

For $\delta b = 0.01$, particles cover only one side of the $\alpha$ hemisphere continuing to travel along $z$
(purple line in Figure~\ref{fig:10va}). This is the transient regime already observed in 
Figure~\ref{overlap_pdf} when $f(\mu)$ exhibits a one-sided long tail distribution toward 
smaller $\mu$. 

For $\delta b = 0.1$, pitch angle distribution becomes completely isotropic and spatial diffusion sets in, as shown
by the slope $s=1$ of $\langle \left( \Delta z \right)^2 \rangle$ in Figure~\ref{fig:10va} (red line) at the end of the simulation. 
The deviation from purely free-streaming or ballistic behavior means that, 
while the system has not become fully diffusive along the mean field direction, there are signs that diffusive processes
in velocity space are beginning to diminish the free-streaming. 
Although $f(\mu)$ still exhibits a long-tail, the influence of spatial diffusion starts to appear. 
The well-pronounced peak observed in $f(\mu)$  for $\delta b = 0.01$ is substantially reduced and
the mean value of magnetic moment decreases. Moreover $\mu$ displays subdiffusive behavior up to $0.02 \tau_c$.
After this time particles diffuse in space and $\langle \left( \Delta \mu \right)^2 \rangle$ attains a plateau. 
The Gaussian shape is not reached yet probably because spatial diffusion is just at the beginning.

\begin{figure*}
\begin{center}
\includegraphics[width=16cm]{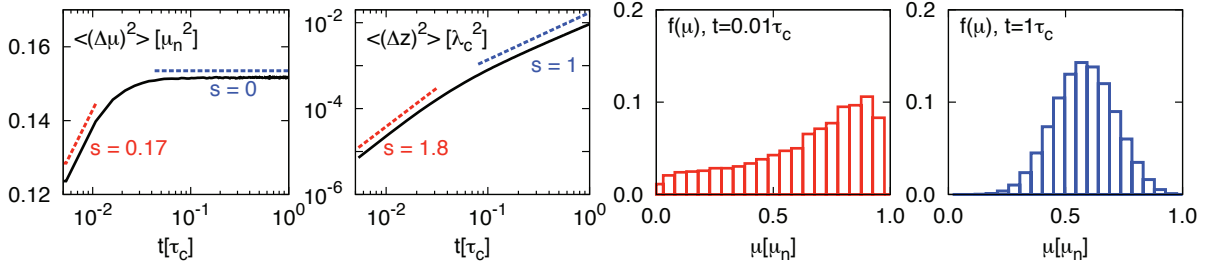}
\caption{Statistics for $v = 10\, v_A$ and $\delta b = 0.5$. Magnetic moment variance $\langle \left( \Delta \mu \right)^2 \rangle$ (first plot) and mean square displacement $\langle \left( \Delta z \right)^2 \rangle$ (second plot) versus time and magnetic moment distribution function $f(\mu)$ after $0.01\, \tau_c$ (third plot) and at the end of the simulation (fourth plot). The $s$ values indicate the different slopes for the 
scaling of the variances $\langle( \Delta \mu )^2 \rangle$ and $\langle( \Delta z )^2 \rangle \propto t^s$. The variances are fitted with the dashed lines.}
\label{deltab_0.5}
\end{center}
\end{figure*}

For $\delta b= 0.5$ [see Figure~(\ref{deltab_0.5})] and $\delta b=1.0$, $f(\alpha)$ is isotropic, particle motion is completely diffusive in real space 
[as the slope $s=1$ in $\langle \left( \Delta z \right)^2 \rangle$ in Figure~\ref{fig:10va} (blue line) shows], 
and $f(\mu)$ behavior is closely related to the sampling of varying magnetic field strength associated with 
that spatial diffusion, displaying a Gaussian distribution centered at the middle of $\mu$-space.

From a more detailed analysis of the case $\delta b = 0.5$ [Figure~(\ref{deltab_0.5})] we notice that magnetic moment variance
(first figure) scales according to $\langle \left( \Delta \mu \right)^2 \rangle \propto t^{0.17}$  (red-dashed line) up to $0.002\tau_c$ and after $0.005\tau_c$ a plateau is attained (blue-dashed line).
Instead particles motion (see time evolution of $\langle \left( \Delta z \right)^2 \rangle$, second plot) becomes fully diffusive (blue-dashed line) only after $0.007\tau_c$. Magnetic moment distribution $f(\mu)$ in the first part of the evolution (third plot) is in the transient regime characterized by the long tail. In contrast, when particles diffuse in real space (fourth plot), $f(\mu)$ reacquires the gaussian profile.
Thus the final stage of magnetic moment variance evolution, i.e. the formation of the plateau, can be considered as a precursor for the onset of the parallel diffusion of particles in space. Of course this effect is present in pitch angle variance too, but in addition in  $\mu$ behavior we have a direct signature of the onset of the spatial diffusion, that is the reappearance of the gaussian shape in the distribution function, while pitch angle distribution remains completely isotropic.

Thus these transitions in magnetic moment behavior are related not just to 
the variation of the turbulence level, but also to the different time scale at which 
magnetic moment conservation is studied.

The magnetic moment distribution functions $f(\mu)$ and variances $\langle \left( \Delta \mu \right)^2 \rangle$ 
in the case $v = 100\, v_A$ (not shown) exhibit the same features observed for $v = 10 v_A$.
However, increasing particle speed the total number of gyroperiods, $N_{\tau_g}$, performed by each particle decreases; as a consequence, faster particles sample less variation in magnetic field strength. This leads to a slower spatial diffusion, i.e., for $100v_A$ spatial diffusion occurs on a time scale longer than $20 \tau_c$. 
\begin{figure}
\begin{center}
\includegraphics[width=8.5cm]{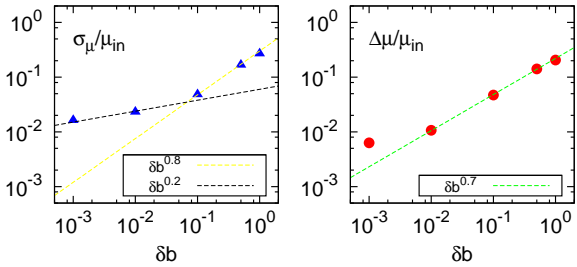}
\caption{Standard deviation (blue triangle) and variation in magnetic moment (red circle)
versus $\delta b$ for at $20 \tau_c$ $v = 100v_A$ and $\alpha = 0.125$.}
\label{sigma}
\end{center}
\end{figure}

For $v = 100v_A$ we show in Figure~(\ref{sigma}) magnetic moment standard deviation $\sigma_\mu/\mu_{min}$ (blue triangle) and the changes in its mean $\Delta \mu/\mu_{in} = (\bar{\mu} -\mu_{in})/\mu_{in}$ versus $\delta b$ after $20 \tau_c$. As $\delta b$ increases toward unity the changes in magnetic moment distribution start to increase faster.

\section{Conclusions}
\label{sec:conclusions}

In this paper we have investigated the conservation of charged particle magnetic moment 
in presence of turbulent magnetic fields. 
For slow spatial and temporal variations of the magnetic field respect to the particle gyroradius and gyroperiod,
the magnetic moment $\mu$
is an adiabatic invariant of the particle motion. Non-conservation of magnetic moment can influence 
particle acceleration and have considerable implications in many astrophysical problems such as 
coronal heating, cosmic rays transport and temperature anisotropies in the solar wind.
These applications motivate the present basic study of the degree to which magnetic moments are conserved 
in increasingly complex models of one dimensional spectra. While all the models considered here have been 
very oversimplified relative to the spectra observed for example in the solar wind \cite{BieberEA94, BieberEA96}
or in simulations of MHD turbulence \cite{OughtonEA94, LeheEA09},
the present study is intended to contribute to the basic understanding of the conditions for the onset of magnetic moment nonconservation.
We point the interested reader also to a recent study by Ref.~\cite{LeheEA09} that addresses this issue from a somewhat different perspective.  

In order to reproduce and extend some of the result obtained by Ref.~\cite{KarimabadiEA92},
we started to study the \emph{ resonant interaction between ions and a single parallel propagating 
electromagnetic wave} (see Section~\ref{sec:onewave}). Using the specialized expression for the 
trapping width $\Delta v_{\parallel}$ found by Ref.~\cite{MaceEA12} in the case of a  single circularly 
polarized wave, we have been able to write a similar expression for magnetic moment (see Eq.~\ref{mu_trap_a}). 
In presence of a single finite amplitude fluctuation the magnetic moment of a resonant particle 
undergoes a finite amplitude nonlinear oscillation too.
We have performed several simulations changing both particle velocity and the amplitude of the wave. 
For each of them we compare the values of $\Delta \mu$ and $\Delta v_{\parallel}$ with those obtained 
using our specialized expression and they are in good agreement.

We designed a particular experiment to study the \emph{effects of resonances overlapping} (see Section~\ref{sec:overlap}).
From the analysis of the distribution functions of particles pitch angle, $f(\alpha)$, magnetic moment, $f(\mu)$, and 
$z$-position, $f(z)$, we distinguish three different regimes.
First, for a low level of magnetic fluctuation, i.e., $\delta B/B_0=0.001, 0.01$, 
the magnetic moment distribution half-width is directly related to pitch angle distribution.
Second for $\delta B/B_0=0.1$ stochasticity arises as a consequence of overlapping resonances and its effect 
on pitch angle is isotropization of the distribution function. This is a transient regime during which 
magnetic moment exhibits a one-sided long-tail distribution and starts to be influenced by
 the onset of spatial parallel diffusion.
Finally, when $f(\alpha)$ completely isotropizes spatial diffusion sets in ( $\delta B/B_0=0.1$), $f(\mu)$ behavior is 
closely related to the sampling of varying magnetic field strength associated with that spatial diffusion.

Other studies regarding particles interaction with two electromagnetic waves as well as a flat 
turbulent spectrum (not shown) were also conducted and they confirmed this general picture. 

Motivated by these results we studied the \emph{behavior of many particles interacting with a broad-band 
slab spectrum}, generated in order to mimic some of the major features of the solar wind (see Figure~\ref{fig:sk}):
(a) three decades of the energy containing scale ensure turbulence homogeneity, (b) three decades of inertial range well-reproduce 
the observations and (c) two decades of  dissipation range enable us to cross the ``$\alpha_{min}$ barrier'' 
related with the ``resonance gap'' predicted by quasilinear theory \cite{KaiserEA73, KaiserEA78}. 
After that there are almost other two decades of zero-padding, important for the trigonometric 
interpolations and for the smoothness of the field. This is implemented using a numerical grid with
$N_{z} = 2^{28} = 268,435,456$ points corresponding to $134,217,728$ wavevectors for the spectrum. 
Apart from the obvious limitation that this spectrum is purely one dimensional, it is constructed to correspond 
roughly to features of solar wind spectra observed by single spacecraft, where the fully three dimensional spectrum
is in effect reduced to a one dimensional form. Information is lost in the process \cite[see, e.g.,][]{BieberEA96}.  

In order to gain insight on magnetic moment conservation we have performed simulations 
changing both particles velocity, $v = (10, 100)\, v_A$, and the amplitude of magnetic field fluctuations 
$\delta B/B_0 = (0.001, 0.01, 0.1, 0.5, 1.0)$. 
Particles  injected at different velocities start to resonate at different 
points of the spectrum. 
We analyzed  the distribution function [see Figure~\ref{fig_10vadistr}] and the variance [see Figure~(\ref{fig:10va})]
of pitch angle cosine $\alpha$, magnetic moment $\mu$ and parallel position $z$.

From the experiment of resonances overlapping we know that the three different regimes of 
$\mu$ statistical behavior are related with other two effects: diffusion in velocity space and spatial parallel diffusion. 
These take place at different characteristic times, $\tau_c$ and $\tau_{\parallel}$ respectively. 
In order to investigate the effects of both processes on magnetic moment distributions, we followed 
test-particles in the simulation box for times $T > \tau_c$.

For a low level of magnetic fluctuations particles free-stream in the $z$-direction 
while $\alpha$ and $\mu$ exhibit gaussian distributions around their initial values.
For $\delta B/B_0 = 0.01$ particles cover completely one side of the $\alpha$ 
hemisphere continuing to stream freely along $z$. 
This is the transient regime during which $f(\mu)$ exhibits a one-sided long tail 
distribution in the direction of smaller $\mu$ that appears to be a typical feature of 
magnetic moment distribution. During this transient regime the distribution of particles 
nearly conserves its magnetic moment. 
Increasing the value of $\delta B/B_0$ spatial diffusion starts to take place, $f(\mu)$ 
recovers the typical Gaussian shape centered in the middle of $\mu$-space.
These different regimes of magnetic moment statistical behavior are related not just to 
the variation of the turbulence level $\delta B/B_0$, but also to the different time scale at which 
magnetic moment conservation is studied [see Figure~(\ref{deltab_0.5})].  

In spite of the limitations of the present approach the results presented here provide a basic view of how magnetic moments are modified in simplified models, and in particular how magnetic moment changes are related to pitch angle changes and sampling of magnetic variations due to spatial diffusion. It is clear that additional study is required to understand more fully the influences of turbulence on magnetic moment statistics. For example realistic three dimensional models of the magnetic field turbulence, as well as incorporation of electric field fluctuation effects, are expected to have significant effects. It is also possible that nonGaussian features of magnetic field fluctuations, such as, are associated with intermittency effects, may also influence magnetic moment changes, much as they influence spatial transport due to trapping and related influences \cite{RuffoloEA03, ChuychaiEA07}.
In this regard the present results, along with those of Ref.~\cite{LeheEA09}, may be considered as baseline or minimal quantification of nonconservation of magnetic moments of a distribution of test particles in turbulence. Planned future studies will investigate quantitatively how additional realism in the modeling might produce even more significant departures from magnetic moment conservation.

\begin{acknowledgments}
This research supported in part by the NASA Heliophysics Theory program NNX11AJ44G, and by the NSF Solar Terrestrial  and SHINE programs. (AGS-1063439 \& AGS-1156094), by the NASA MMS and Solar probe PLus Projects, and by Marie Curie Project FP7 PIRSES-2010-269297 - Turboplasmas.
\end{acknowledgments}
 
\appendix*
\section{Derivation of trapping half width for a circularly polarized wave}

Using equations $(5a)$ and $(5b)$ of \cite{KarimabadiEA92} 
it is possible to derive a simplified expression for the trapping half-width 
and the bounce frequency in the case of an Alfv\'en static wave \cite{MaceEA}. 
For this particular case $k_{\perp}=0$ and $\phi=0$. 
We can rewrite equation~$(5c)$ of \cite{KarimabadiEA92} as
\begin{align}\label{zn1}   
Z_n &=  mc^2 \left\{ \frac{v_{\perp}}{2c} \left[ \left( \epsilon_2 - \frac{k_{\parallel}}{k}\sigma \epsilon_1 \right) J_{n-1}(k_{\perp \rho}) + \right. \right. \nonumber \\
       &\left. -  \left( \epsilon_2 + \frac{k_{\parallel}}{k} \sigma \epsilon_1 \right) J_{n+1}(k_{\perp \rho}) \right] + \nonumber \\
       &\left. +  \sigma \left( \frac{v_{\parallel} k_{\perp}}{ck} \epsilon_1 + \epsilon_3 \right) J_n(k_{\perp \rho}) \right\},
\end{align}
with $\cos\alpha=1$ and $\sin\alpha=0$. Because ${\bf k} \parallel {\bf B}_0$ we can choose 
${\bf \hat{e}}_z={\bf B}_0/|{\bf B}_0|$, ${\bf \hat{e}}_y$ is any arbitrary direction perpendicular 
to ${\bf \hat{e}}_z$ and ${\bf \hat{e}}_x= {\bf \hat{e}}_u \times {\bf \hat{e}}_z$. 
The vector potential can be obtained from the magnetic field
$\nabla \times {\bf B}_{\perp} = B_x {\bf \hat{e}}_x + B_y {\bf \hat{e}}_y$. 
In Fourier space $\nabla \rightarrow ik_z {\bf \hat{e}}_z$, so we have:
\begin{eqnarray}\label{potential}
A_x & = & - \frac{i}{k_{\parallel}}B_y \nonumber \\
A_y & = &  \frac{i}{k_{\parallel}}B_x
\end{eqnarray}
Considering only a single circularly polarized wave in space, for the two different possible helicities
we can write:
\begin{equation}
{\bf B}_{\pm}  = (B_\pm {\bf \hat{e}}_\pm) \exp{[i(k_{\parallel}z)]}
\end{equation}
where
\begin{equation} 
B_\pm  = \frac{1}{\sqrt{2}}(B_x \mp iBy) \quad \textrm{and} \quad
{\bf \hat{e}}_\pm =   \frac{1}{\sqrt{2}}({\bf \hat{e}_x} \mp i{\bf \hat{e}}_y)
\end{equation}
are respectively the complex amplitudes and the orthogonal polarization unit vectors.
The $+ (-)$ polarization state is the positive (negative) helicity, i.e., the vector ${\bf B}$ 
is rotating counter-clockwise (clockwise). 
At first, let's consider only the left-handed polarized wave ${\bf B}_+$. 
Assuming $B_+ = \sqrt{2} \delta B e^{-i \pi/2}$ we can write the $x$ and $y$ components 
of the wave magnetic field as
\begin{eqnarray}\label{bwave}
B_x & = & \delta B \exp{[i(k_\parallel z - \pi/2)]} \nonumber \\
B_y & = & \delta B \exp{(ik_\parallel z)}
\end{eqnarray}
Inserting this two expressions into Eq.~\ref{potential} we obtain:
\begin{eqnarray*}
A_x & = & \frac{\delta B}{k_\parallel} \exp{[i(k_\parallel z - \pi/2)]} \nonumber \\
A_y &  =  &\frac{\delta B}{k_\parallel} \exp{(ik_\parallel z)}
\end{eqnarray*}
Comparing the real parts of these equations with equation~$(1b)$ of \cite{KarimabadiEA92} 
we obtain an expression for the coefficients $A_1$ and $A_2$ and for the normalized 
components of the wave polarization vector $\epsilon_1$, $\epsilon_2$ and $\epsilon_3$:
\begin{equation}\label{potential1}
A_1 =  \eta\frac{\delta B}{k_\parallel},     \quad     A_2 =  \frac{\delta B}{k_\parallel},      
\quad  \textrm{where}   \quad      \eta=\frac{k_\parallel}{|k_\parallel|}
\end{equation}
\begin{equation}\label{potential12}
\epsilon_1 = \frac{|q|\eta \delta B}{mc^2 k_\parallel},        \quad        \epsilon_2  =  \frac{|q|\delta B}{mc^2 k_\parallel}, 
\quad    \epsilon_3=0.
\end{equation}
Similarly, for a right-handed circularly polarized wave ${\bf B}_-$ we have:
\begin{equation*}\label{potential2}
A_1  =  - \eta\frac{\delta B}{k_\parallel},     \quad     A_2 =  \frac{\delta B}{k_\parallel},
\quad  \textrm{where}   \quad      \eta=\frac{k_\parallel}{|k_\parallel|}
\end{equation*}
\begin{equation*}
\epsilon_1  =  - \frac{|q|\eta \delta B}{mc^2 k_\parallel},        \quad        \epsilon_2  =  \frac{|q|\delta B}{mc^2 k_\parallel},
\quad    \epsilon_3=0.
\end{equation*}
In case of a single circularly polarized wave propagating parallel (or antiparallel) to the magnetic field 
there is only one resonance present and particle motion is integrable \cite{KarimabadiEA92}: 
indeed $J_n(0)= 0$ unless $n=0$. Therefore depending on the polarization of 
the wave and on its direction of propagation $\eta$ only $l=1$ or $l=-1$ resonances contribute 
to the trapping width, as shown in Table~\ref{tab_trapping}.
\begin{table}
\caption{Wave polarization and resonance contribution to trapping width.}
\center
\begin{tabular}{ | l | c | c | }
\hline   
Polarization                          &    $\eta$                      &      resonance $ n$  \\
\hline
$B_+$ left-handed               &    $1$ parallel              &       $-1$                   \\
$B_-$ right-handed              &    $-1$ anti-parallel      &       $1$                    \\
$B_+$ left-handed               &    $-1$ parallel              &       $1$                    \\
$B_-$ right-handed              &    $1$ parallel               &       $-1$                   \\
\hline
\end{tabular}
\label{tab_trapping}
\end{table}
Thus, considering  equations $(5a)$ and $(5b)$ of \cite{KarimabadiEA92}, Eq.~\ref{zn1} and 
Eq.~\ref{potential1}-\ref{potential12} with $J_0(0)=1$, \cite{MaceEA} find a 
specialized formula for the trapping half width and bounce frequency applied to the case of 
a circularly polarized wave propagating parallel $k_\parallel > 0$ and $n=-1$, 
or antiparallel, $k_\parallel > 0$ and $n=1$ to ${\bf B}_0$:
\begin{eqnarray}\label{deltav}
\Delta  {v_\parallel}^{(-1)} & = & 2v \left[(1-\alpha^2)^{1/2}|\alpha|\frac{\delta B}{B_0}\right]^{1/2} \nonumber \\
{\omega_b}^{(-1)} & = & \Omega_0\left[\frac{(1-\alpha^2)^{1/2}}{|\alpha|}\frac{\delta B}{B_0}\right]^{1/2}
\end{eqnarray}
if $k_\parallel v_\parallel > 0$ and zero otherwise, in which $\alpha=\cos\theta$ is the cosine of pitch angle. 
Exactly the same set of equations holds for $\Delta v_\parallel^{(+1)}$ and $\omega_b^{(+1)}$. 
However the condition for their being nonzero is reversed, i.e., $k_\parallel v_\parallel > 0$. 
We omit the superscripts $(\pm1)$ because of this degeneracy.

\clearpage

\end{document}